\documentclass[a4paper,12pt]{llncs}
\usepackage[top=4cm,bottom=4cm,left=3.9cm,right=3.8cm]{geometry}

\usepackage[T1]{fontenc}

\usepackage{floatflt}
\usepackage[tight]{subfigure}
\usepackage{latexsym}
\usepackage{clrscode}
\usepackage{natbib}

\usepackage{ifpdf}
\ifpdf 
 \usepackage[pdftex]{graphicx}
 \else
  \usepackage{graphicx}
\fi
\usepackage{listings}

\author{Thomas C. Schmidt\inst{1} \and  Arne Hildebrand \inst{2} Michael Engelhardt \inst{3} \and Dagmar Lange \inst{3}}
\institute{HAW Hamburg Department Informatik,
 Berliner Tor 7, D--20099 Hamburg, Germany
\and
link-lab, H{\"o}nower Str. 35, D--10318 Berlin, Germany
\and
HTW Berlin, Hochschulrechenzentrum. Treskowallee
 8, D--10318 Berlin, Germany 
\\\email{t.schmidt@ieee.org, info@link-lab.net, \{engelh,langeda\}@fhtw-berlin.de}}

\title{From a Link Semantic to Semantic Links --- Building Context in Educational Hypermedia}

\begin{document}
\maketitle

\begin{abstract}
Modularization and granulation are key concepts in educational content management, whereas teaching, learning and understanding require a discourse within thematic contexts. Even though hyperlinks and semantically typed references provide the context building blocks of hypermedia systems, elaborate concepts to derive, manage and propagate such relations between content objects are not around at present. Based on Semantic Web standards, this paper makes several contributions to content enrichment. Work starts from  harvesting multimedia annotations in class-room recordings, and proceeds to deriving a dense educational semantic net between eLearning Objects decorated with extended LOM relations. Special focus is drawn on the processing of recorded speech and on an Ontological Evaluation Layer that autonomously derives meaningful inter-object relations. Further on, a semantic representation of hyperlinks is developed and elaborated to the concept of semantic link contexts,  an approach to manage a coherent rhetoric of linking. 

These solutions have been implemented in the Hypermedia Learning Objects System (hylOs), our  eLearning content management system. hylOs is built upon the more general Media Information Repository (MIR) and the MIR adaptive context linking environment (MIRaCLE), its linking extension. MIR is an open system supporting the standards XML and JNDI. hylOs benefits from configurable information structures, sophisticated access logic and high-level authoring tools like the WYSIWYG XML editor and its Instructional Designer. 
\end{abstract}
\paragraph{Keywords:}  Educational Content Management, Semantic Web, Link Context, Educational Semantic Net, eLearning Objects, Metadata, Keyword Spotting, Machine Reasoning
\setcounter{footnote}{0}
\section{Introduction}

The Semantic Web was created as an initiative to bring machine processable structure to the bulk of Web information, which initially had been designed for human reception only. A vision was presented of robots and crawlers that digest online material on a  level of understanding sufficient to create high-level mediators \citep{bhl-tsw-01}. Positioned as a new kind of human-machine interfaces, intelligent agents were proposed to act between content and recipients. First and foremost they should provide superior navigational knowledge to facilitate instant information retrieval. 

Navigational intelligence, in contrast, has been a subject of thorough investigations in the community of open hypermedia systems. The focus of long-term research has been on modeling and designing uncomplex systems of sufficient openness and conceptual freedom to serve the needs of rich application appearance. Numerous Web frameworks have been created with that aim and in particular the XML languages for online document processing. The field of hypermedia has been freshly inspired by manifold open and distance learning activities. Numerous domain-specific applications have been created, as well as standardized meta description frameworks. In the context of systematic knowledge acquisition and learning, hyper-referential functions and content classification directly interrelate with approaches to instructional design.

Combining the fields of semantic and hypermedia opens the perspective on human-centric hyper-systems that are grounded on high-level intelligence generated by a network-transparent, machine-processable semantic information layer. The realm emerges for applications  which exceed the scope of a search machine, but  offer a richer vision of a semantically assisted human-machine interaction. Navigational intelligence should take a central design role in this growing area, as it not only forms the core component of interactivity in any hypermedia application, but also carries the major contribution of context between portions of online content. 
A user following a flow of reception from pieces of online content is bound to the paths offered by links. He should experience hyperreferences  that are meaningful to the content itself and to his perspective of perception. Here, links are considered essential parts of the content and should follow a rhetoric as characterized in the early work of \cite{l-rhm-89}. In dynamically operating content management it is therefore important to autonomously identify and the thematic context, as well as the perspective of the user.

Since several years, a growing number of application-specific communities is concerned about metadata descriptions of their information bases, acting in parallel to the semantic web initiative. Their motivations do not only derive from subject-specific categorization and retrieval, but also from the challenge of automated content processing and context-centric presentation.  In this work and the examples given below, we concentrate  on the agile area of educational content management and related metadata descriptors.  A corresponding standard for the annotation of educational material has been released with the IEEE Learning Object Metadata \citep{ieee-lom-02}. 

In the field of educational content management, the concept of atomic, self-consistent content units was derived from the paradigm of object-oriented modeling and became known as eLearning Objects (eLOs) cf., \cite{w-lodst-00}. It has been widely accepted as a standard entity in combination with LOM metadata. eLOs may combine rich media content, a significant set of metadata and structural relations. eLO content itself may be composed of other eLearning objects, thereby constructing a self-similar knowledge tree suitable for navigation in this fashion.  The LOM metadata subsume technical, textual and educational information, which are complemented by a set of named relations. Originally, the latter had been defined for technical and administrative purposes, but shall prove to be valuable for  constructing a mesh of inter-object guidance to the learner. LOM has been chosen as part of the high-level exchange format SCORM, the  \cite{adl-scorm}.

In this work, we discuss semantic approaches to  interactivity in  educational hypermedia systems, dedicating special focus to building context from inter-object relations. Concepts and implementations are based on our eLearning content management system hylOs, which adheres to a strict separation of content structure, logic, data and design by applying XML-technologies in a rigorous fashion. 
Our approach starts from metadata extraction of multimedia content, covers a semantic model of annotated links that are grouped in link contexts, and arrives at an automated reasoning process for generating dense semantic nets between loosely coupled eLearning objects. We will discuss a general semantic representation of hyperlinks and the adaptive linking environment MIRaCLE, a  model and implementation of a semantic link processor along the lines of this article.
We also formalize the semantic of an extended set of eLO relations within an ontology and supply an additional set of inference rules. Starting from some initial relations, any new object entering a repository can then automatically harvest named links to any other object from the concurrent processing of an inference engine. Initial relations can automatically be derived from previous automatic classification and annotation, manual provision of hyperlinks or from a limited manual relation setting.

The remainder of his paper is organized as follows. In section \ref{sec:hylos} we  briefly introduce the Hypermedia Learning Object System hylOs, our eLearning content management platform used for all implementations. We further on present  our design and implementation of  an automated  eLO content acquisition and speech analysis in section \ref{sec:acquisition}. Section \ref{ref:linking} derives the concept of a semantic hyperlink model, as well as the MIRaCLE scheme for processing  contextual information of hyperlinks.
Section \ref{section:oel} proposes semantic specifications and extensions to the LOM relations and discusses an ontology-based inference processing. Finally, section \ref{sec:c+o} is dedicated to a conclusion and an outlook on future work.

\section{The Hypermedia Learning Object System hylOs}
\label{sec:hylos}

\begin{figure}
\begin{center}
\includegraphics[width=\columnwidth]{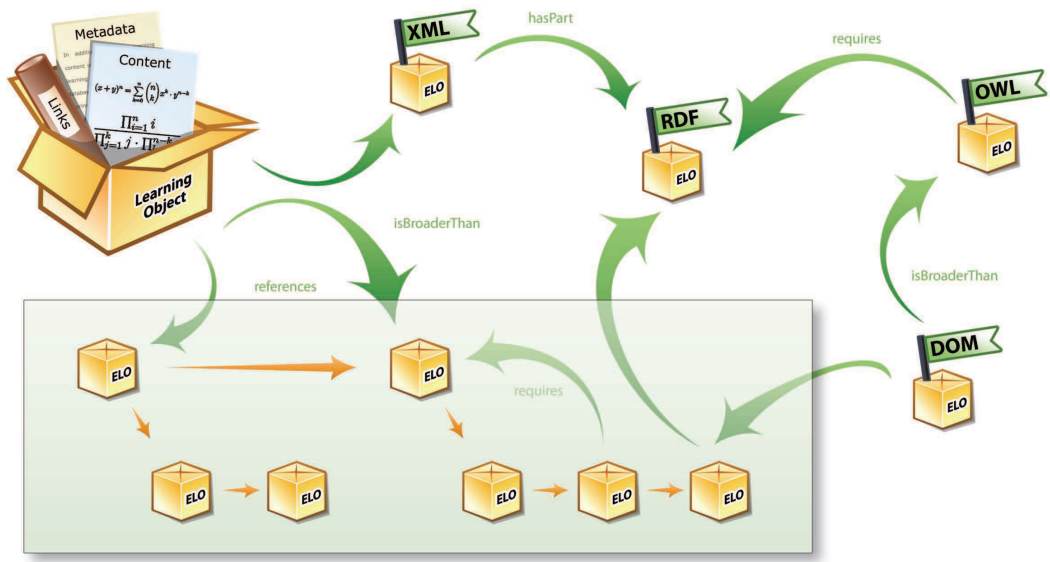}
  \caption{eLearning Objects in Semantic and Instructional Nets Offer Variable Content Access}
  \label{figure:elo-net}
\end{center}
\end{figure}

The Hypermedia Learning Object System \citep{ehkrs-ecmca-02,se-ecm-05,hylos} has been designed to provide full educational content management based on the eLO information model. All knowledge units are composed of rich media content elements decorated with a complete set of IEEE LOM metadata and interconnected by qualified relational pointers. They reside within the Media Information Repository \citep{fs-motms-01,fkrs-aepcm-01}. Grounded on a powerful media object model, MIR was created as a universal fundament for ease in modelling and implementation of complex multimedia applications. All data embedded within the adaptable MIR data store are published in XML format, such that individual views and specific user interfaces can be produced from lightweight style sheets. The rigorous use of the XML technology framework ensures a consistent separation of content, structural information, application logic and design elements. hylOs' adaptive eLearning functions are thereof derived and may attain any look \& feel by applying appropriate XSL transforms. The system is used in several eLearning and content management deployment projects within our institutions and commercial training environments. 

hylOs offers variable content access views to the learner. Figure \ref{figure:elo-net} visualizes the different content network layers that can serve as a dynamic foundation for user navigation:  Aside from the primary content tree elaborated by the author(s), instructional design hierarchies may be compiled from repository objects for each teaching trail. Based on its qualified relations, the content additionally is organised in a semantic net, suitable for individual exploration in a constructivist fashion as will be described in detail in section \ref{section:oel}. Traditional hyperreferences, which provide a separate layer of content traversal, may be customized within hylOs, as well. 

hylOs treats links as first class content objects in a highly dynamic fashion. To consistently execute content linking, the MIR Adaptive Context Linking Environment MIRaCLE \citep{es-cbaih-03}  processes links as semantic statements, its more general concepts will be discussed in section \ref{ref:linking}.  Distinguished hyperlink layers may be applied onto the same content, as may have been predefined by the teacher or selected by the learner. Links are represented within contextual containers, each one suitable to express a  narrative of a specific hyperlinking scheme. These link contexts may be understood as a composition of link rhetoric, as suggested in the early work of  \cite{l-rhm-89}. Note that textually coherent hyperlink collections provide an additional, meaningful structure to be harvested in future applications.

A fully distributed authoring environment is part of the hylOs suite, as well. Authors are enabled to edit eLOs in full detail, i.e., rich media content (including mathematical formulae), the LOM metadata tree and all types of relations. Great care has been taken to simplify content elaboration wherever possible. An 'easy authoring sheet' within the SWING application provides WYSIWYG XML editing combined with extensive automated harvesting of metadata. Manual provision for only seven LOM attributes  are needed, i.e., keywords, semantic density, difficulty, context, learning resource type, structure and document status, if presets taken from previous editing do not apply. While creating subsequent eLO content, authors implicitly generate an object tree.  Re--use of content and structures is supported at any level of complexity. A variety of specific editors for glossaries, (\LaTeX--compliant) bibliographies and taxonomies complement this first--level authoring suite.

Assisted by an additional authoring sheet, the Instructional Designer  \mbox{(iDesigner)}, any instructor will be enabled to compose overlay trees individually designed for a specific teaching trail. The process of forming a didactically structured outline from single learning components is commonly known as instructional design. This task of arranging the eLOs in different courses or units remains outside the scope of the eLO paradigm. 
Our approach to instructional design introduces the idea of an instructional container object (ICO). ICOs are inherited from eLOs retaining the complete LOM metadata set and the nesting facility. They implement either a structural container for nesting other ICOs or a visual container to embed eLOs. In addition they offer appropriate instructional types and an optional prolog or epilog. Instructional types are courses, sections or pages, for example.
In this way, the hylOs iDesigner bridges the gap between re-usability and atomicity of eLearning objects on the one hand, and individually and coherently designed courses on the other. It very flexibly imposes instructional overlays onto any, possibly loose collection of eLearning objects.

Several, often commercial systems for authors are around to edit and generate SCORM packages. Few of them address issues of editing complex content structures and exhaustive, semi-automated metadata support. A carefully prepared hypermedia editing environment, the hypermedia composer HyCo, has been introduced by \cite{gg-ehrf-05}. HyCo aims to seamlessly guide authors, while provisioning structural and semantic information for educational content objects. In taking such content-oriented, non-technical perspective, the system is closest to our hylOs approach.

\section{From Content Acquisition to Automated Classification}
\label{sec:acquisition}

\subsection{eLO Acquisition}
A manual preparation of eLearning objects remains a tedious undertaking, no matter how well it is supported by appropriate tools. In addition, most presentation material currently in use is brought to lecture rooms by notebooks or offline media, not compliant to the LOM/IMS packaging standard. The goal of our automatic content acquisition subsystem is therefore to produce eLOs 'out of the lecture room'. Complying to a predefined quality standard, these objects may then be used for rapid play out or manual refinement. We start from  observing that audiovisual streams are available for capturing within modern, A/V--equipped lecture halls and concentrate on visual presentation material in combination with spoken audio. We abstain from taking video recordings of the lecturer in our standard scenarios, as shooting and play--back of video require high efforts, while adding limited value to the learning.

\begin{figure}[h]
\begin{center}
\includegraphics[angle=270,scale=0.14]{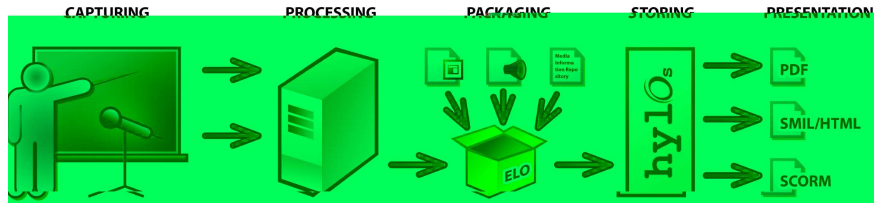}
  \caption{Process Flow of the Content Acquisition Subsystem}
  \label{figure:processing}
\end{center}
\end{figure}

 The signals, i.e., images and audio streams, are continuously captured and  analysed. For confining suitable sections, the perpetual media stream is segmented following triggers, which are fired at a change of slide or presentation material. Appropriate cuts are derived from the audio by a silence analysis: Within a neighbouring interval of $\pm 10~s$  of the trigger moment, a large pause, $ \geq 0,5~s$, of the speaker is identified and used for the cut. Pauses are detected according to dynamically evaluated water\-marks. Still in real--time, the obtained audio segment will undergo a rescaling of volume and pause lengths to approach a sound even in volume and rhythm of speech. Finally the data will be efficiently compressed using the Ogg Speex codec \cite{speex-codec}.  

After segmentation the appropriately encoded media will be packaged and -- annotated with technical metadata -- stored as raw eLOs within hylOs as shown in figure \ref{figure:processing}. Satisfying the full LOM data structure, the eLearning objects obtained so far are suitable for rough online consumption, manual refinement or further automated processing.

\subsection{Automated Classification}

The raw eLOs do not contain any semantically valuable metadata nor do they admit qualified relations to the remaining repository. Subsequent post processing targets at deriving suitable multimedia annotations \citep{sops-masw-06}. First and foremost, a keyword assignment will be needed as well as a classification according to a predefined taxonomy to the newly acquired objects. To achieve this goal, all available sources of information are used, which consist of a predefined context, a  predefined classification system in particular, text written on slides and the spoken words from the audio. However, for practical reasons we cannot presuppose a training phase on any speech recognition system and have to refrain from the use of complete speech to text transcriptions. Thus little textual material is available and pure statistical techniques cannot be used. 

Instead, we employ a controlled vocabulary for keyword spotting. A selection of several thousand pre--classified technical terms is searched within the text and audio files. Speech recognition is done using the Philips Speech SDK\footnote{Philips Speech SDK: 
http://www.speechrecognition.philips.com/}. This SDK offers the option to dynamically activate user--specific vocabularies defined in the Java Speech Grammar Format \citep{jsgf-98}. JSGF allows for the generation of tokens as a combination of synonyms, which we use for collecting flexion forms. Flexions of German words have been harvested from a Web Service of the project "Deutscher Wortschatz" at Leipzig\footnote{Project "Deutscher Wortschatz"
at Universit\"at Leipzig: http://wortschatz.uni-leipzig.de/Webservices/}.

After keywords have been extracted, we further proceed in a dictionary based approach to derive the classification indices of the eLearning Object by using a simplified version of the statistical schemes derived by \cite{r-ir79} and \cite{mza-acsadls01}.  Operating only on keyword sets, regardless of document lengths, classification is done according to a keyword hit rate for each taxonomic node.  We employ Jaccard's coefficient in our heuristic approach and vary significance levels around $0.5$. Note that this procedure  is not based on a rigorous statistical approach, but follows an experimentally driven optimization.  Current classification is done with respect to  \cite{ddc} and \cite{acm-ccs}, the approach generalises, though, without restrictions.  
By analyzing text and speech content the recorded eLearning objects has thus been enriched by a title, the author, keywords and classification categories. Additional educational attributes such as context or age range may be harvested from presets within the session or from profiles associated with the author.

In the following section \ref{section:oel} we will take up these results to derive semantic relations between eLOs from the annotations and classifications so far obtained.

\subsection{Evaluation}

We emphasize  in advance that the procedure introduced above must be considered as a semi--heuristic approach to generate semantic metadata for raw content. There is no guarantee for completeness or correctness, which needs to be assured by manual post--editing, whenever desired. Nevertheless, we consider it a noticeable advancement to seamlessly retrieve multimedia content objects decorated with expressive, but semi-reliable metadata, rather than to insist on authors provisioning, which is rarely completed and may be mistaken, as well (see \cite{ps-siqf-04,nktvmd-ueloi-04} for a detailed discussion).

In detail, the approach of keyword spotting has proven to work reasonably well with the untrained speech recognition systems, provided the employed number of terms remains small. For optimisation, we partition our dictionary into sets of up to 50 keywords, selected according to an iteratively narrowing context along the classification hierarchy. Term sets then are subsequently applied. This procedure assures not only a largely improved reliability in keyword identification, but allows for consideration of pre--set context branches in taxonomic hierarchies and scales logarithmically.  Fixing a {\em precision} of 75 \%, an average {\em Recall} of about 80 \% is obtained. Failures dominantly originate from unrecognised keywords, while false positives remain rare. 
On a current generation of server system results indicate that calculation efforts stay within real-time bounds.\footnote{We use a standard PC with a multicore 3 GHz processor. Note that keyword splitting allows for parallel processing of different term sets.}

The selection and evaluation of keywords we consider the crucial part. As there seem to be no classified terminologies for the German language publicly available, we extracted a collection of nearly 2.000 technical terms from bilingual English--German encyclopaedias, which we complemented by additional sources and domain-specific knowledge. Classification of these key--term collections has  not only been tedious work, but a reliable source of inexactness. Consequently the quality and richness of our classified vocabulary varies with respect to the subfields visited in the ACM CCS. Major improvements may be achieved in future refinements. 

Classification results do not arrive at a uniform level of exactness: A  spotted keyword "Ethernet" for example will easily lead to highly reliable branches of the ACM taxonomy, whereas solely identified terms like "System" or "Design" will not. Presently the largest shortcoming of our scheme must be seen in its limited ability to arrive at statistically significant classifications, the latter naturally being a function of the preset significance level.

\section{Metadata, Semantic and Hyper-Relations}
\label{ref:linking}
\subsection{Identifying the Semantic of Hyperlinks}
\label{sec:link-sem}
\subsubsection{Metadata and Anchors}

The common approach of the Semantic Web targets at provisioning resource descriptors and ontologies to robots such as search machines. Consequently, facing the current status of information in the Web, a major effort concentrates on the acquisition of semantic statements describing resources and on building up ontological databases. Prior and in parallel to these  activities, the hypermedia research community has completed more than 40 years of research concerned with the organisation and interrelation of content \citep{ohr-hswra-02} components, focusing concepts and development on an advanced design of network--transparent hyperlinks.

\begin{figure}[th]
\begin{center}
\includegraphics[width=14cm]{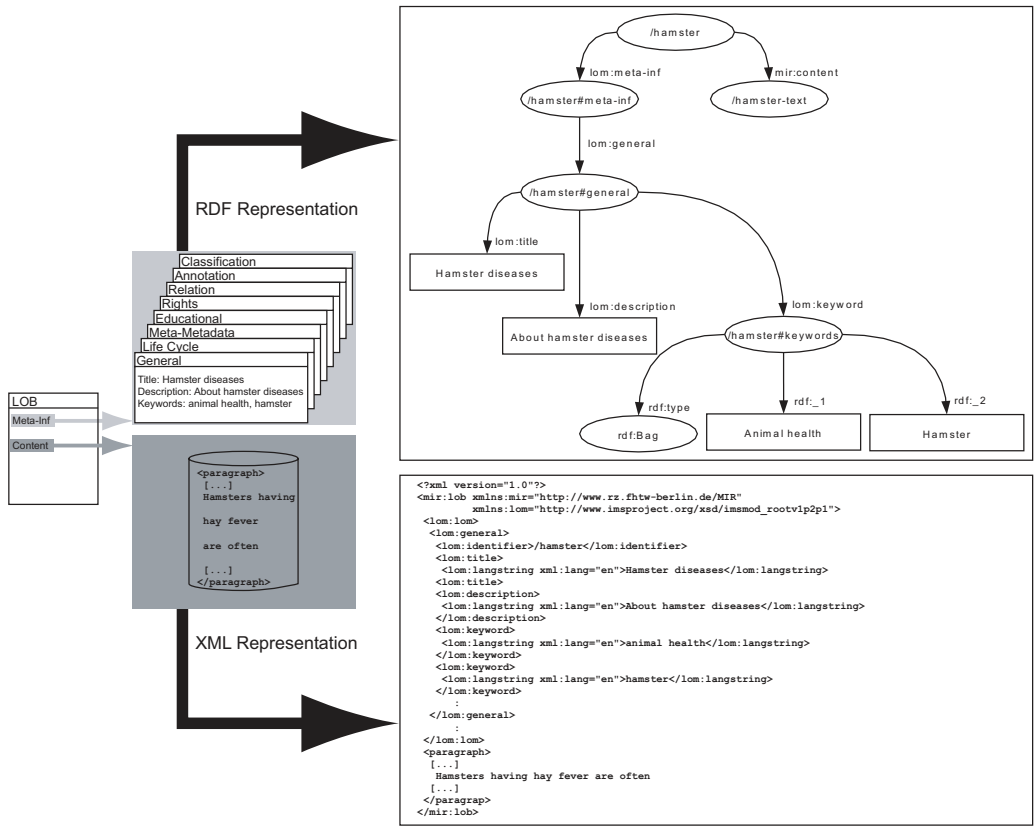}
\caption{Representation of Metadata in XML and RDF Statements}
\label{fig:meta-rdf}
\end{center}
\end{figure}

In the previous sections, we have discussed the nature and semi-automated derivation of IEEE LOM metadata.  
Assuming those metadata in presence now, a canonical semantic description of content items can be easily derived: In an  \cite{w3c-rdf} representation, the content object attains the role of the subject, the name of the meta descriptor forms the predicate, and its value denotes the object. Figure \ref{fig:meta-rdf} illustrates the different representations of the statement ``This learning object is a description about hamster diseases.'' following an LOM/XML schema and RDF graph notation.

To approach a semantic analysis of hyperlinks, let us recall that a hyper-reference is constructed of two entities, anchors and links. Links connect anchors which identify sub-portions of content. In a fairly general fashion, anchors can be formulated within XLink \citep{w3c-xlink-01} statements by XPointer/XPath-like expressions \citep{w3c-xpath-07,w3c-xpointer-03}, the exact formalism depending on the media type of the document. Links, as well as anchors may be stored separated from document resources, e.g., in a link  base.

\begin{figure}[tb]
\begin{center}
\includegraphics[width=\textwidth]{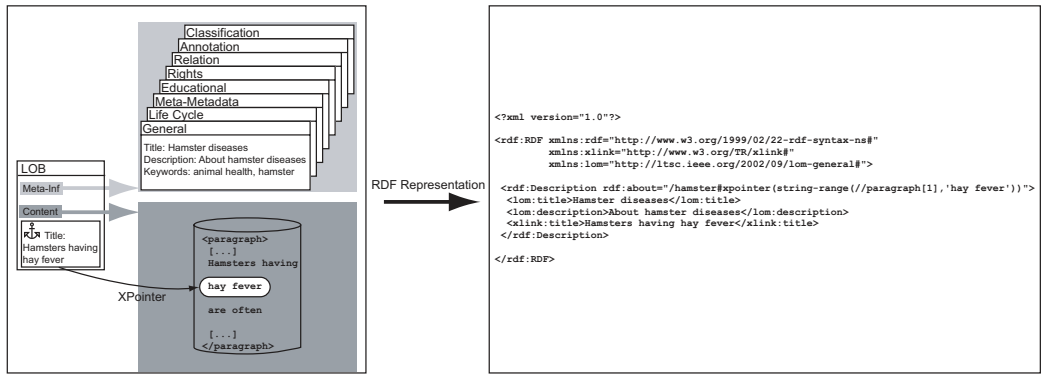}
\caption{Harvesting Additional Anchor Descriptors from Metadata}
\label{fig:meta-anchor}
\end{center}
\end{figure}

Even though it may appear straightforward to inherit a semantic description of an anchor from the expository statements of the associated content object, simple information inheritance remains insufficient. This limitation becomes evident for a document that carries several anchors at fragments of distinct meaning. It is therefore important to provide additional specifications which can be harvested from the title and label tags of XLink locator expressions.\footnote{Annotations may also appear in `Resource' statements embedded within the content. Here a conceptual deficit of the Xlink specification becomes apparent: While using a link base with locators, attributes denoting anchored resources cannot be assigned outside actual link definitions.} Note that the denoted data chunks in anchors need not be of textual type. Anchors in this sense must be viewed as additional specialisations, e.g., ``This resource in the context of hamster diseases carries the title `Hamsters Having Hay Fever'''. The extraction of a semantic description of anchored resources given as a collection of inherited and dedicated statements is visualised in figure \ref{fig:meta-anchor}.

\subsubsection{Link Semantic}

Links denote relations between two or more anchors. They are directional components, i.e., uni- or bi-directional and bear the immediate expressiveness of (contextless) relations: ``Resource A is pointered to B''. Following the XLink arc encoding, a link expression itself may carry directionless attributes such as multiple titles, as well as directional descriptors like the arc attributes 'from', 'to' and 'arcrole'. 
A contextual semantic of hyperlinks will account for these meta descriptors and match attributes, e.g., using arcroles whenever arc and direction apply. 
Thereby, link attributes give rise to a collection of simple statements: ``This link carries the title `For Freshman''', ``This link starts from the resource `Hamsters Having Hay Fever''', etc..

\begin{figure}
\begin{center}
\includegraphics[width=\textwidth]{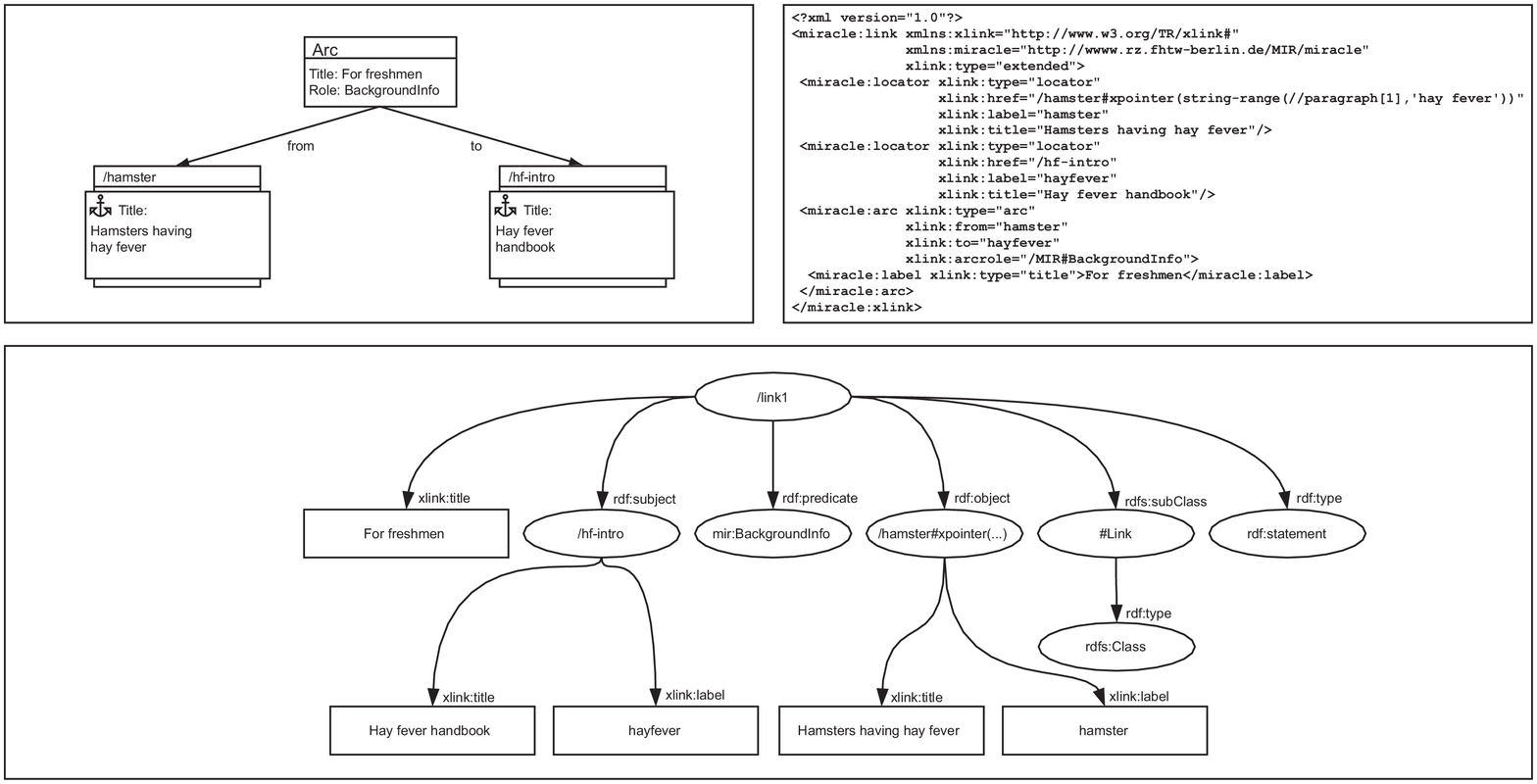}
\caption{From XLink to an RDF Hyperlink Description}
\label{fig:xlink-rdf}
\end{center}
\end{figure}

 In semantic terms, statements represent  resources that are linked, and a link encodes a relation between them. The link itself carries additional meaning in attributes. Thus an XLink expression gives rise to a more complex, reifying or second order statement.  Via its arcrole attribute, a link expresses a predicate describing the  resources it refers to. An example of such a link meta descriptor is visualized in the upper part of figure \ref{fig:xlink-rdf}, which characterizes the linking relation as a ``BackgroundInfo'' reference.

Transforming this notion into a simple statement, e.g., ``Resource `Hay Fever Handbook' presents BackgroundInfo to resource `Hamsters Having Hay Fever''' will cause a loss of context, since the link resource itself will not be part of the statement representation. To cure this deficit, a higher order statement, a statement about statements needs to be employed. Following this line of thought, the link entity forms the subject of a statement about the statement that describes the relation. As visualised in figure \ref{fig:xlink-rdf}, such expression reads, ``Link1 denotes that resource `Hay Fever Handbook' presents BackgroundInfo to resource `Hamsters Having Hay Fever'''

Expressing the core semantics of hyperlinks as higher order statements opens the opportunity to address the link entity itself as a first class object. It preserve the relation to context information such as link titles, authors, additional attributes etc.. Viewing the approach in a rigorous semantic fashion, this is indeed correct, as a link may form a resource external to content, and denote a relation not valid by itself. Rather a link serves as an expression of the perspective and potentially personal view of its (link) author, who may be distinguished from content authors. 

\subsection{An Application Scenario}

Interactivity plays a dominant role in educational learning systems. Well-organised content can be significantly enriched or disturbed by the way links add interactivity to it. As discussed in the introduction, we do not consider the definition of links as part of the content itself, but rather as part of the didactic structuring and presentation model. In particular, we foster ways to stir the process of link selection and presentation when applied  to the  content.

To illustrate this argument in a simple use case, let us return to our previous example: A short introductory overview on hamster diseases written in the Gaelic language is presented to a Scottish veterinarian, who has significant weakness in Gaelic. The utmost help for his understanding will probably come with links that reference every word to the corresponding entry in a dictionary. An Irish hamster farmer with some semi-expert knowledge about his animals instead, would profit most from having the medical terms linked to some 	colloquial background explanation. An Irish student of vet learning for his exam  could rather appreciate all disease names being appropriately connected to some collected encyclopaedic knowledge of sample exam questions. And so on.

Reflecting on this example, we can extract the following requirements for an appropriate linking environment:
\begin{itemize}
\item There should be the capability of applying different linking schemes to the same content; thus the definition of links cannot be part of the content itself.
\item Linking should adapt to the users requirements.
\item A user should be enabled to shape the linking within an application according to his requirements.
\item A common narrative meaning of the links should be in place and transparent to the user.
\item High-level mechanisms for defining and selecting links are needed in order to keep work of the author simple.
\item The meaning of link entities should be open for inspection by external content processors.
\end{itemize}
Some of the above requirements can be resolved with the help of Xlink/Xpath/Xpointer, but major issues remain unsolved. Introducing a semantic notion of hyperlinks raises information to a level, where authors and automated link processors within applications may be enabled to place or perform instructions effective and consistent in form and content. In the following section  we will report on our concepts in this regard.

\subsection{Semantic Link Context}

\label{sec:link-context}

The concept of open Hypermedia, i.e., the decoration of content with anchors and links collected from a distinguished link base, gives rise to remarkable flexibility in attaching hyperreferences to hypermedia documents. It is thereby not only possible to apply different linking schemes to the same content, but also to dynamically present link selections adaptively chosen to meet user needs. In applying dynamic linking, though, an application should follow demands on consistency and coherence, a rhetoric pattern for example as characterized in the early, fundamental work of \cite{l-rhm-89} for static link creation.
 Having derived a semantic notion of anchors and hyperlinks in section \ref{sec:link-sem}, we are now able to define a high-level scheme for collecting and processing links from a link-base.
		
		In hypermedia processing the context is an important concept. In a document-centric view, there are already  different contexts to recognise: The context, a document appears or is to be presented in, the context of document fragments, given by its surrounding document data, and the (fragmental) source and destination context of a hyperreference. Here we are concerned with the semantic context of a  link itself. Link contexts are capable of articulating certain orthogonal information such as object-related notions of an author, or a perspective of link creation and traversal, which expresses an inter-resource dependence imposed by the offer to jump between content objects. Both  paradigms are preserved by the semantic link encoding introduced in the previous section.

To exploit these additional encodings, a high-level semantic selection layer is needed to perform operations on link selections and collections based on the link context. Providing such mechanisms will enable authors and audience to steer hyperlink appearance by semantic criteria and thus allows for more precise and purposeful interaction within a hypermedia application. There are many potential operations like extracting links according to their semantic role, attributes or their relationship with underlying anchors. Most importantly, links can be adapted to users within personalised hypermedia applications.
		
		The concept of a link context layer introduces a new abstraction on link collections. Within this layer, semantic rules  for processing links  can be predefined that  dynamically operate on an abstract data model provided by the link layer. Link contexts neither create new links, nor new anchors. They are only responsible for the extraction of existing links stored in a link base. Those links are characterized by their descriptive properties as shown in the previous section. They subsume a selection scheme to represent a desired semantic context.
		
\begin{figure}
\begin{center}
\includegraphics[scale=0.6]{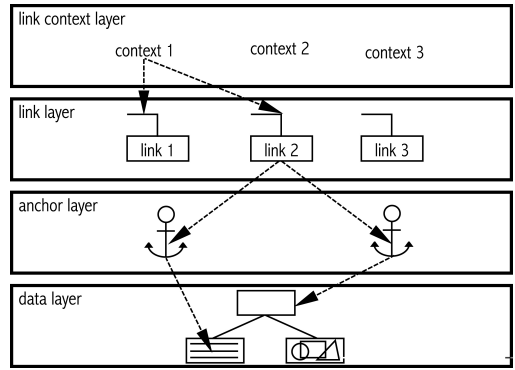}
\caption{The MIRaCLE Link Layer Model}
\label{fig:miracle-layers}
\end{center}
\end{figure}
		Link contexts are the upper tier in a four layered model named MIRaCLE, cf. figure \ref{fig:miracle-layers}, consisting of a data, an anchor, a linking and the link context layer. The MIR adaptive context linking environment (MIRaCLE) \citep{es-cbaih-03} is both, a formal model and a practical implementation based on the MIR system. It results in a ``semantic filtering'' to obtain the appropriate subset of links as requested by a user or an application. All entities, anchors, links and the activated link contexts are processed on the fly while content is accessed  via a standard Web browser.
		
		As all semantically relevant notions from the link or anchoring layer are expressible in a formal RDF model, the link context itself is operating on the model of the link layer representation and enables users to select groups of semantically related links. Retrieving links means extracting sub-graphs from the model. The extraction of sub-graphs is done by an appropriate query language, like SPARQL \citep{w3c-sparql-08}.
		
		The result of such a query are statements, which have the chosen links as subject and at least one predicate-object pair formed by the involved anchors and their relationship. Identifying the subject of the return statements as a link yields all  information required for further processing. Based thereon, an application  can extract the participating anchors, verify them for being a start resource regarding the current document and visualize them as desired.
		
		\begin{lstlisting}[frame=bt, caption=An Example of a Link Context Definition, label=src:sparql-sample]{}
<?xml version="1.0"?>
<rdf:RDF xmlns:rdf="http://www.w3.org/1999/02/22-rdf-syntax-ns#"
         xmlns:mir="http://www.rz.fhtw-berlin.de/MIR"
         xmlns:dc="http://purl.org/dc/elements/1.1/">
<rdf:Description rdf:about="link-context1">
  <dc:Creator>Rainer Zufall</dc:Creator>
  <dc:Title xmL:lang="en">Background Information</dc:Title>
  <dc:Description xml:lang="en">
    Some continuative information on.
  </dc:Description>
  <mir:link-context>
<![CDATA[
	PREFIX rdf:<http://www.w3.org/1999/02/22-rdf-syntax-ns#>
	PREFIX mir:<http://www.rz.fhtw-berlin.de/MIR/mir#>
	SELECT ?subject
	WHERE {?subject rdf:predicate mir:BackgroundInfo}
]]>
  </mir:link-context>
</rdf:Description>
		\end{lstlisting}

		Returning to our previous example of a vet student reading the text about hamster diseases, let us imagine a context meant to  provide some background information on the current topic  ``Hamsters having hay fever''. One possible link context definition is given by listing \ref{src:sparql-sample}.

The query will return all matching nodes in the graph which are the subjects of the associated RDF statements. The subjects will contain the name of the appropriate link and a statement about the connected anchors.
In  terms of our example it will return the link which expresses: ``Link1 denotes that resource `Hay Fever Handbook' presents BackgroundInfo to resource `Hamsters Having Hay Fever'''. This higher order statement contains a simple statement embedding the target anchor as the subject, the predicate being the relation and the source anchor the object. There is all necessary information for rendering the link into the document.

\begin{figure}
\begin{center}
\includegraphics[width=10cm]{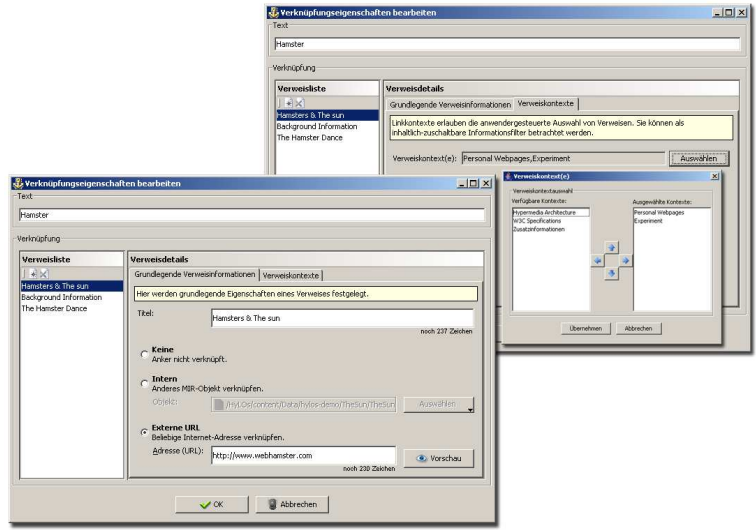}
\caption{Authoring Environment for Editing Links in Contexts}
\label{fig:link-editor}
\end{center}
\end{figure}

While the generality of this model may cause worry about authoring complexity, it can actually be transformed into simple, end-user compliant operations: Once parameterizable SPARQL code, which forms the backend for link context definitions, is  provided in advance, link contexts can be defined by setting terms. An author is then just obliged  to place link meta values and select a context. A corresponding authoring interface from the hylOs system is shown in figure \ref{fig:link-editor}.

\subsection{Related Work}
Activities of joint research in a 'semantic hypermedia' have been identified, see \cite{ohr-hswra-02} for an excellent overview. Our work tackles the open problems  of a semantically driven hyperlink service stated therein and addresses the fundamental question of dealing with possible inconsistencies of knowledge, when fragments of distributed knowledge resources are hyperlinked: By encoding hyperlinks as reified statements, hyper-relations are viewed within their individual context. These contexts will be encoded by the author of a link following his or her perspective. Thus inconsistencies will arise only if contradictory link statements of compatible context are given, which then can be traced back to their origins. 

Initially not starting from  a semantic perspective, the importance of context for link processing has been emphasized by \cite{ehrc-lic-02}. The work of the Southampton group traditionally ranks around virtual links and anchors, which are subject to a run-time computation at the reception time. The basic idea presented lies in a dynamical derivation of multi-destined links for (Web) resources, as extracted from a link base. This work has been extended to an ontology-based approach for link generation in order to ``escape from the limitations of a purely lexical string matching'', cf. \cite{chbg-clobo-01,maklhds-gahcsw-03}. From the hypermedia perspective, these concepts are closest to our approach. In contrast to our work presented here, the group attempts to determine a link context by analyzing the anchor context and to prevent misleading links by ambiguous phrases. Their runtime environment is built on a multi agent architecture performing distributed tasks synchronised via a message passing model. This realisation completely diverges from our data-centric approach.

The gap between the syntax oriented XLink and the semantically oriented RDF is bridged in the W3C note  \cite{w3c-hrsx-00}. This work focuses on the extraction of semantic statements from XLink metadata. Since both, XLink and RDF, are capable of describing relations between resources, it is quite natural to 'harvest' semantic expressions from hyperlink encodings. According to the note, the arcrole attribute as the most significant metadata for characterizing the relation of participated resources is expanded to an RDF statement with the start resource as the subject, the end resource as the object and the value of the arcrole attribute denoting the predicate. Following our previous line of argument, this approach does not formulate statements about links, but sole statements on anchors. Problems and shortcomings of this document centred view has been discussed above.

Several recent projects attempt to vitalize the visions of the Semantic Web on  grounds of resource tagging provided by Web users. The Simile project with the MIT Piggy Bank Web application by \cite{hmk-pbesw-07}  is one of the prominent examples. The main objective of these structurally and semantically open Web initiatives lies in a broad user involvement. Initiators aim to accumulate meaningful Web annotations, which can be shared and shall enable a semantically assisted smart browsing. Tagging within the Simile project follows a property:value pair mapping and gives rise to statements of the kind ``Resource A is in a relation to B, characterized by <property:value>''. It cannot be concluded, wether the characteristic map denotes the link annotation itself, i.e., ``This tag is for testing.'', or the link traversal, i.e., ``See this for background info.'', which is the major shortcoming of this fully unrestricted approach.

\section{An Ontology Based Approach to Constructing Educational Semantic Nets} 
\label{section:oel}
\subsection{Advancing LOM Relations}
\label{section:lom-extend}

eLearning objects compliant to the LOM metadata standard provide a section of qualified object relations that allows to interconnect any two objects in a meaningful fashion.  Assuming a well maintained mesh of eLOs in place, a semantic learning net may be presented to the learner for navigation and knowledge exploration, as well as to the author or instructional designer.  

However, the expressiveness of LOM relations is limited to the administrative view of librarians, as types and semantics of these relations have been directly adopted from the Dublin Core library metadata set (see table \ref{table:lom-relation-properties}). To gain expressions suitable for educational hypermedia, the syntax of DC relations requires a careful extension, while its semantic needs adaptation  and sharpening.

\begin{table}[h]
	\centering
	\begin{tabular}{|l|l|l|l|l|l|}
		\hline
		\small{Is part of} & \small{Is version of} & \small{Is format of} & \small{Is referenced by} & \small{Is based on} &
		\small{Is required by}\\
		\hline
		\small{Has part} &  \small{Has version} & \small{Has format} & \small{References} & \small{Is basis for} & \small{Requires}\\
		\hline
	\end{tabular}
	\caption{Original Dublin Core/LOM Relations}
	\label{table:lom-relation-properties}
\end{table}
 
To design a fairly comprehensive, viable notion on a semantic educational net, we proceed in three phases: At first, we select those relations from the DC set which remain suitable under minor modifications and specifications in the educational hypermedia context. The results are shown in table \ref{table:modified-relation-properties}. The major, unobvious change consists in turning 'isFormatOf' into a symmetric property. The corresponding DC inverse property pair expresses bibliophilic editorial hierarchies, which remain absent in hypermedia systems.

At second, we redefine the semantic of those DC properties that originally had been bound to pure technical terms. Even though similar re-interpretations have been commonly undertaken in LOM based educational contexts, an explicitly stated semantic is lacking, but needed for further operations. Table \ref{table:redefined-relation-properties} displays the corresponding entities and their semantic definitions. These three DC property pairs now essentially express thematic dependencies of increasing strengths. 'references' and 'isBasedOn' both express an optional value to the learner, and thus cannot pursue transitivity.

\begin{table}[h]
	\centering
		\begin{tabular}{| p{4cm}|p{8cm}|}
			\hline
			Relation & Semantic \\
			\hline
			\hline
			hasPart/isPartOf & This inverse pair of transitive properties expresses the structural relation of nesting eLOs. There is no additional meaning related to content.\\
			\hline
			hasVersion/isVersion & This pair of inverse properties describes versioning as generated by updates or redesigns. Different versions may deviate in content and author, preserving thematic dedication and technical format, though.\\
			\hline
			isFormatOf & This symmetric property relates eLOs which essentially cover the same content in different formats. It does not imply interchangeability, but a persistence of educational context.\\
			
			\hline
		\end{tabular}
		\caption{Modified Semantic for Selected DC Relations}
	\label{table:modified-relation-properties}
\end{table}

\begin{table}[ht]
	\centering
		\begin{tabular}{| p{4cm}|p{8cm}|}
			\hline
			Relation & Semantic \\
			\hline
			\hline
			references/ \mbox{isReferencedBy} & This inverse property pair describes a weak form of content relation: An author references another eLO for extended information -- similar to a common use of hyperlinks.\\
			\hline
			isBasedOn/  \mbox{isBasisFor} & This inverse property pair relates an eLO carrying content fundamental to another. It expresses a semantically strong, but optional thematic relation.\\
			\hline
			requires/  \mbox{isRequiredBy} & This inverse pair of transitive properties denotes a mandatory content dependence in the sense that eLO A cannot be understood without knowledge of eLO B.\\
			\hline
		\end{tabular}
		\caption{Redefined Semantic for Selected DC Relations}
	\label{table:redefined-relation-properties}
\end{table}

Finally we design a small set of additional relation properties which are missing in the LOM standard. Most importantly, the taxonomic interdependence 'isBroaderThan' has been raised to the eLO net. Guided by the maxim of restraint, only three horizontal relations have been introduced to improve orientation in content access. 
Any additional terms expressing a meta discourse on content, as introduced by \cite{ssfs-mmw-99}, were omitted for the sake of simplicity and clarity. Types and semantic of these newly introduced relations are summarized in table \ref{table:new-relation-properties}.

Note that all relations occur symmetric or in inverse pairs. Besides systematic considerations, this characteristic covers an important technical advantage. Any author may add any relation simply to his own objects, without requiring write access to the corresponding pair.

\begin{table}[t]
	\centering
		\begin{tabular}{| p{4cm}|p{8cm}|}
			\hline
			Relation & Semantic \\
			\hline
			\hline
			isNarrowerThan/  \mbox{isBroaderThan} & This inverse pair of transitive properties encodes the standard taxonomic relation.\\
			\hline
			isAlternativeTo & This symmetric transitive property connects interchangeable eLOs. Alternative eLOs are meant to be of equivalent content, pedagogical and structural properties, but may deviate in formats.\\
			\hline
			illustrates/  \mbox{IsIllustratedBy} & This inverse property pair expresses illustration in an open fashion. For illustration an eLO need not be of specific content type. \\
			\hline
			isLessSpecificThan/  \mbox{isMoreSpecificThan} & This inverse pair of transitive properties relates two objects, which are of strong thematic familiarity, but differ in generality. A more specific object may cover sub-aspects or the identical subject in more detail or exhibit a thematic overlap while being more specific.\\
			\hline
		\end{tabular}
		\caption{Additional Educational Relations}
	\label{table:new-relation-properties}
\end{table}

A visual  example of a semantic net derived from extended LOM relations is shown in figure \ref{figure:mindmap}. All subjects displayed from the Programming Language context are connected via qualified relations that allow for a coherent, semantically guided content access. 
A learner, facing a well maintained educational semantic net tied by the relations described above, will greatly profit in orientation, content navigation and exploration. It is moreover easy to implement, as has been done within the hylOs application. Any author exploring the current state of an eLO repository will likewise benefit from a dense thematic overview of his region of interest.

\subsection{Ontological Evaluation Layer}

 Adding a new eLearning object will require to identify and update appropriate relations with a possibly large amount of repository entries. Objects entering the repository by automated acquisition as described in section \ref{sec:acquisition}, will be predisposed as unconnected entities. A manual maintenance of large content networks soon becomes infeasible in any realistic setting.  

Classified objects, though, offer sufficient information for a continued automated processing. They may immediately inherit the relations 'isBroaderThan' from the taxonomy and 'isPartOf' from their structural disposition. Additional attributes may be conjectured from heuristic considerations, e.g., two eLOs of (almost) identical classification and keyword sets, as well as comparable educational attributes are likely to be 'AlternativeTo' each other.   Inserting the objects "HTML" or "Markup\_Languages" for example can initiate a fully automated generation of named interconnects.

To overcome the obstacle of further manual netting, an {\em Ontological Evaluation Layer (OEL)} \citep{ehls-raemo-06} has been designed and implemented in hylOs. The core concept consists of encoding relation semantics within an  \cite{w3c-owl-04} ontology, which then can be processed by an inference engine. At this first step, relation pairs along with formalized characteristics can be distributed across a repository. To account for logical dependencies between related properties, additional inference rules need to be supplied to the inference engine. As outcome of a careful overlook, we identified about 50 of such rules, giving rise to a dense inference set. Some typical examples of inherent conclusions read: 

\begin{itemize}
  \item $A$ is narrower than $B \wedge B$ is format of $C \Longrightarrow  A$ is narrower than $C$
  \item $A$ is based on $B \wedge C$ has part $B \Longrightarrow A$ is based
  on $C$
  \item $A$ requires $B \wedge B$ is based on $C \Longrightarrow A$ is based on
  $C$
  \item $A$ is more specific than $B \wedge B$ is format of $C \Longrightarrow A$ is more specific than
  $C$
  \item $(A$ is version of $B \vee A$ has version $B) \wedge A$ is format of $C \wedge B$ is format of $C$ \newline $ \Longrightarrow A$ is alternative to $B$
\end{itemize}

Our implementation uses the JENA framework \citep{JENA} to execute the reasoning, combining the extended relation ontology and the additional inference rule. A daemon triggered by object insertion or update within the repository concurrently adds appropriate relations to the new or changed object. By following a strategy of concurrent evaluations that immediately become persistent in the repository, our hylOs implementation accounts for the rather slow reasoning process of the JENA framework, which is unsuitable for real--time interactivity.  

\begin{figure}[ht]
\begin{center}
\includegraphics[width=\textwidth]{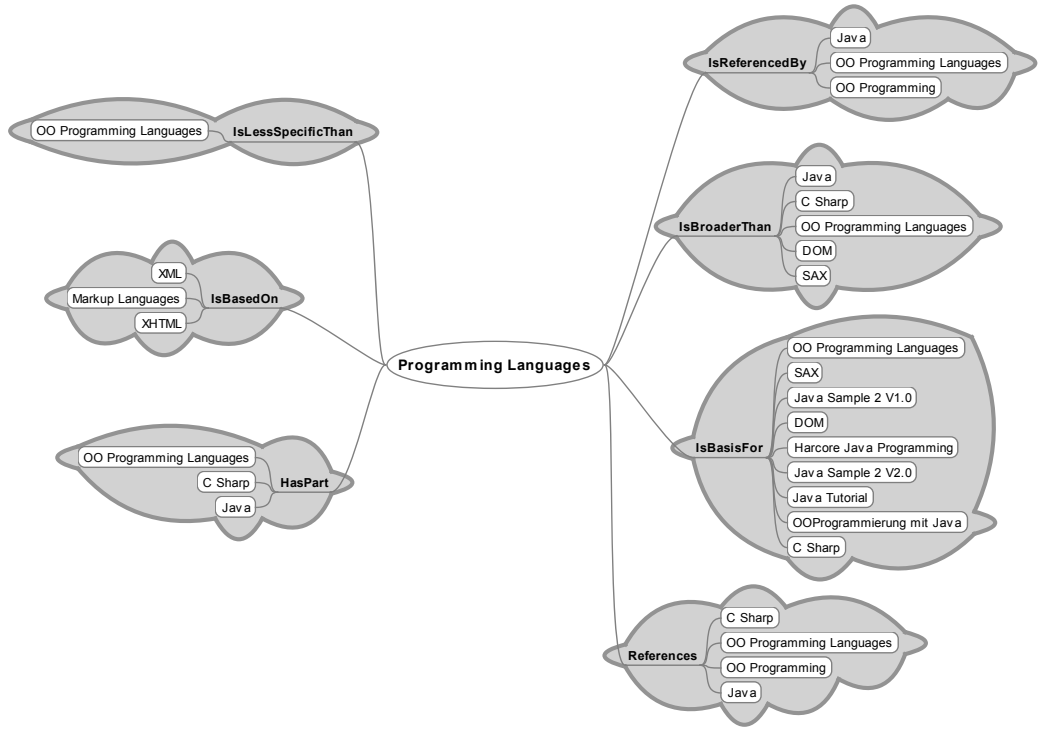}
  \caption{Application Example: Mindmap Navigation}
  \label{figure:mindmap}
\end{center}
\end{figure}

\begin{figure}[t]
\begin{center}
\includegraphics[width=\textwidth]{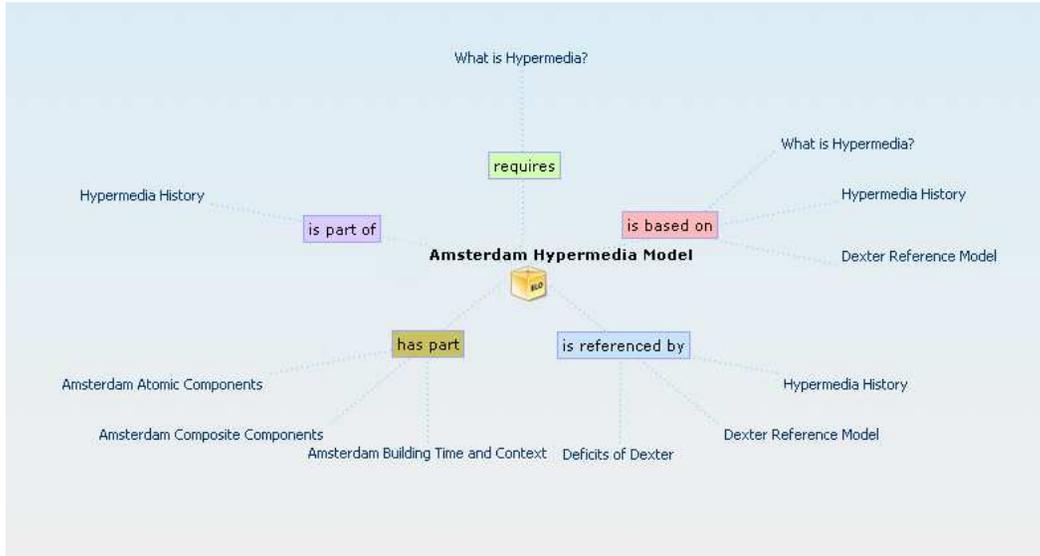}
  \caption{Application Example: Navigational Net}
  \label{figure:navnet}
\end{center}
\end{figure}

 Any newly inserted object or relation  will thereby lead to a chain of subsequent link placements within the hylOs system.  Authoring thus is enriched by a forceful augmentation intelligence. Learners will profit from automated reasoning and envision a consistent and supposedly dense educational semantic net. One of manifold application opportunities built on top of this semantic net is displayed in figure \ref{figure:mindmap}. An eLO--centric mind map type of view is dynamically generated from all relations of the currently displayed content, offering hyperlink navigation according to the semantic net perspective. Another interactive view of the relational mesh is shown in figure \ref{figure:navnet}. This Web 2.0 AJAX layer is used as a visual overlay for navigation in one of our front-ends deployed for current learners. 

Such content navigation schemes may equally be offered to learners, instructors and authors. While an author may intuitively explore content related to his personal editing, an instructor may be guided to specific content areas, e.g., for supplementary instructional use. Finally, a learner will experience a 'looking glass' onto his study field, giving rise to a self-directed constructivist-kind of knowledge access. Such a view bares a transparent perspective not only on personal study needs, but also on conceptual interdependencies of subject areas.

\subsection{Evaluations}

For a thorough evaluation of the procedures and the axiomatic rule set introduced above, we proceed in two ways.  At first, we analyse efficiency of the proposed rigorous and heuristic schemes, i.e., a quantification of the gain in relations obtained by automated reasoning. At second, we design and apply a semantic monitor to identify inconsistencies and incorrect relations that have been generated by our automatic reasoning process. This crosscheck analysis yields an a posteriori, experimental validation of the inference rules and procedures used in semantic reasoning.

For the quantitative analysis, we select sample collections of eLOs from real--world teaching and proceed in the following way. We initially apply rigorous  reasoning on the object set, which is structured and classified, but bare of any preset relation, and obtain result set $\cal{A}$. In the second step, we add conservative heuristic conclusions, from which few relations of the types 'isAlternativeTo',  'isVersionOf' and 'isMoreSpecificThan' derive and subsequently apply full reasoning to receive result set $\cal{B}$. At this stage, we assume a typical contribution of an author who sparsely added thematic dependencies, which will be processed by the reasoner into result set  $\cal{C}$. Finally, we manually add major relations that we identified as missing and receive result set $\cal{D}$ after a last processing run. 

\begin{figure}[t]
\begin{center}
\includegraphics[width=\textwidth]{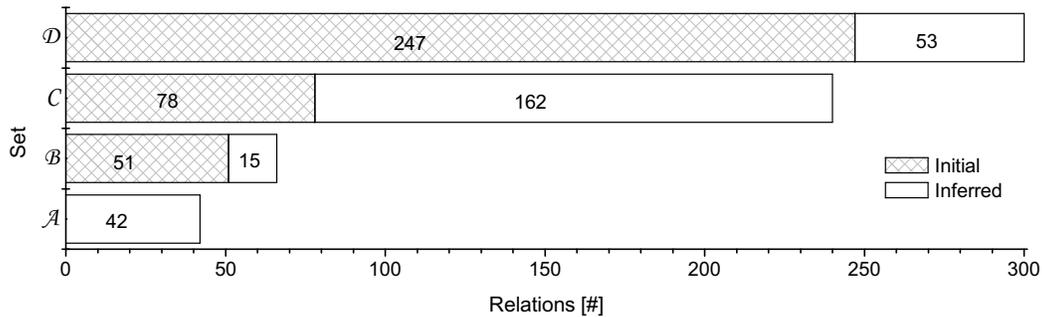}
  \caption{Yield of Inferred Relations for Different Initial Conditions}
  \label{figure:reasoner-results}
\end{center}
\end{figure}

Results for a sample set of 18 eLOs consisting of the two thematic islands "OO Programming" and "XML Technologies", exhibiting "DOM" and "SAX" as points of contact, are displayed in figure \ref{figure:reasoner-results}. In transition from set $\cal{A}$ to $\cal{B}$ 9 relations have been added by heuristics. 12 were produced by a virtual author in changing to $\cal{C}$ and 7  misses have been contributed to initiate result set $\cal{D}$.

Overall it could be observed that a dense mesh of 300 relations has been created in this procedure, where 66 have been derived from \'a priori and heuristic conclusions. 206 semantic links were descended from 28 manual addings. Even though many of them stem from a direct transfer of classification or were concluded by a few straightforward steps, they may carry value by linking formerly  unrelated resources. 
Aside from quantity, it should be noted that the inferred relations are not at all limited to the obvious; according to the underlying effective rule set, quite surprising results occur. The two initially unrelated eLOs "Markup Languages" and "Java Sample 2" inherit for example the relation "isRequiredBy" through the following chain: "Markup Languages" $\stackrel{hasPart}{\longmapsto}$ "XML", "DOM" $\stackrel{requires}{\longmapsto}$ "XML" and finally "Java Sample 2" $\stackrel{hasPart}{\longmapsto}$ "DOM".

This quantitative experiment demonstrates the effectiveness of the rule-based reasoning process, which proved to produce a densely interwoven mesh of content relations. All content objects have been interconnected with  17 link an average. These semantic relations provide thematic context and coherence to a loose collection of eLearning content objects. Note that relations need not be utilised in a raw manner, but may be projected to a reduced quantity or complexity, whenever users or devices require. A corresponding example for mobile devices is presented by \cite{hse-mecd-07}, where relations have been adapted to the navigation interfaces of the iPOD and a Portable Sony Playstation. 

We will now proceed with crosschecking the reasoning system. A proof of correctness for the proposed rules requires a multistage analysis and is only achievable up to the semantic precision inherent in eLearning content and metadata definitions. Most importantly, evidence is required that our axiomatic rule set is contradiction--free. A contradiction occurs, if two objects are related by mutually contradicting relations. Such contradictions obviously derive from inverse relation pairs, but also from mutually exclusive semantics of unpaired relations such as 'isFormatOf' and 'isVersion'\footnote{'isFormatOf' denotes a change of format, while requiring persistent content, whereas 'isVersion' relates objects of developing content, but excludes changes in format.}. It is worth noting that such semantic exclusions cannot be modeled in OWL directly. To still monitor inconsistencies, which also could result from manual editing, we encode incorrectness--relations in a separate ontology, e.g.,
\begin{itemize}
  \item $A$ is part of $B \wedge B$ is part of $A \Longrightarrow  A$ incorrectPart $B$
  \item $A$ is format of $B \wedge B$ has version $A \Longrightarrow A$ incorrectFormatVersion $B$
\end{itemize}
and apply the reasoner likewise. Discovering contradictory relations in this way then requires a manual root cause analysis based on tracking the corresponding inference chains.

The semantic soundness of the rule set we analysed in two ways. Theoretically we carefully evaluated each single rule with respect to all formal semantic aspects of the related eLOs. Empirically we crosschecked many collections of "real--world" eLearning Objects, decorated with all relations we could accept as true. Subsequent automatic reasoning was applied to each collection and any additional result was taken up for an individual examination. In an early stage of our work, this experimental monitoring proved helpful to identify inconsistent inference rules. After a conservative consolidation, our inference system has reached a stable balance in the sense that incorrect semantic links are only produced from incorrect object classifications or failures in manual editing. Thereby, it is our steady experience that erroneous inferences are quickly discoverable, since the outcome of rule-reasoning rapidly spreads among entities and soon leads to obvious incorrectness. 
Further on a continuous monitoring of inconsistency can be concurrently applied and will then establish a firm, current judgement on the correctness of any deployed content net. 

\subsection{Related Work}

 \cite{p-uarlo-03} discusses the relevance of eLearning Object relations to enable a discourse on the information exposed. 
Several authors have addressed details on the employment and enhancement of the IEEE LOM relations \citep{ssfs-mmw-99,vc-acmvc-03,ks-lomlca-04} for educational content. Their work ranges from a more technical oriented course production perspective up to a formal encoding of domain specific knowledge within relations.

The use of semantic meta data for browsing purposes was introduced by \cite{fl-sb-03}. Departing from a librarian perspective of organising information, the authors  pursue a realisation of the early vision of 'associative indexing' by Vannevar Bush \cite{b-awmt-45}.  Not bound to the educational context, \cite{zjl-slnbi-04} define an abstract model of semantic links for facilitating a ``Knowledge Grid''. Their approach  allows for user-defined tagging and linking at the one hand, and a reasoner-based autonomous link building process on the other. To accomplish the latter, the authors derive a rule-set appropriate for evolving their semantic relations in a procedure, which is conceptually comparable to our work.

\section{Conclusions and Outlook}
\label{sec:c+o}

In this paper we discussed key aspects of automated eLearning object acquisition and contextual content augmentation. Starting from the LOM-based Hypermedia Learning Object System hylOs, we initially  introduced our approaches of eLO generation, classification and meta data harvesting from lecture recordings. 

Focussing on context within educational applications, a detailed derivation of hyperlink semantic hase been worked out. By encoding references as second order statements, the defining entity of the link could be preserved within a one-statement representation. A high-level link context layer, which allows for learner-specific selections of hyperreferences, could be gained as an immediate benefit for decorative linking schemes.

 Widening the perspective to inter-object references, an evaluation of the LOM semantic relations was presented, and these technical metadata have been elaborated into an improved relation set. It was shown that by turning the inherent relational logic into operational reasoning, a semantic learning net will actively evolve and monitor its consistency. The resulting inter-object relations give rise to a rich variety of semantically guided content exploration for learners, as well as for authors. A manual provision of such complex structures would be completely infeasible, as authors had to overlook entire repositories to identify the semantic links. 

This approach of generating a semantic content net can be extended to inter--repository overlays in a straightforward way. Our future work will  concentrate on the design of semantic peer--to--peer networks formed  between distributed eLearning Objects, and on its integration in mobile and social network environments.

\section*{Acknowledgment} 

This work has been supported in part by the German Bundesministerium f\"ur Bildung und Forschung (BMBF) within the project Mindstone \linebreak (http://www.mindstone.org).

\bibliographystyle{agsm}
\bibliography{/bib/rfcs,/bib/ids,/bib/internet,/bib/layer2,/bib/overlay,/bib/own,/bib/theory,/bib/transport,/bib/vcoip,/bib/hypermedia}
\end{document}